%% file: main.tex
\documentclass[journal,comsoc]{IEEEtran}

\usepackage{graphicx}
\usepackage{color}
\usepackage{cite}
\usepackage{algorithm}
\usepackage{algorithmic}
\usepackage{url}
\usepackage{subfigure}
\usepackage{bm}
\usepackage{optidef}
\usepackage{amsmath}
\usepackage{amssymb}
\usepackage{amsfonts}

\newcommand{\relmiddle}[1]{\mathrel{}\middle#1\mathrel{}}

\begin{document}
\title{Bayesian Optimization Framework for Channel Simulation-Based Base Station Placement and Transmission Power Design}

\author{
    Koya~Sato and Katsuya Suto~\IEEEmembership{Members,~IEEE}
    \thanks{\\
    This work was supported in part by JST ACT-X Grant Number JPMJAX21AA, JST PRESTO Grant Number JPMJPR23P3, and JST ASPIRE Grand Number JPMJAP2346, Japan.
    Koya Sato is with the Artificial Intelligence eXploration Research Center, The University of Electro-Communications, 182-8585, Tokyo, Japan (e-mail: k\_sato@ieee.org).
    Katsuya Suto is with the Graduate School of Informatics and Engineering, The University of Electro-Communications, Chofu 182-8585, Japan (e-mail: k.suto@uec.ac.jp).
    Corresponding author: \it{Koya Sato}
    }
}
\markboth{ACCEPTED FOR PUBLICATION IN IEEE NETWORKING LETTERS (DOI: 10.1109/LNET.2024.3469175)}%
{K. Sato and K. Suto: Bayesian Optimization Framework for Channel Simulation-Based Base Station Placement and Transmission Power Design}
\maketitle

\begin{abstract}
  This study proposes an adaptive experimental design framework for a channel-simulation-based base station (BS) design that supports the joint optimization of transmission power and placement. We consider a system in which multiple transmitters provide wireless services over a shared frequency band. Our objective is to maximize the average throughput within an area of interest. System operators can design the system configurations prior to deployment by iterating them through channel simulations and updating the parameters. However, accurate channel simulations are computationally expensive; therefore, it is preferable to configure the system using a limited number of simulation iterations. We develop a solver for the problem based on Bayesian optimization (BO), a black-box optimization method. The numerical results demonstrate that our proposed framework can achieve 18-22\% higher throughput performance than conventional placement and power optimization strategies.
\end{abstract}


\IEEEpeerreviewmaketitle

\input{introduction.tex}
\vspace{-2mm}
\input{systemmodel}
\vspace{-2mm}
\input{bo}
\vspace{-2mm}
\input{bo_proposed}
\vspace{-2mm}
\input{performance}

\section{Conclusion}
\label{sec:conclusion}
We proposed a BO framework for channel simulation-based joint placement and transmission power design.
The results demonstrate that our proposed framework can achieve 18-22\% higher throughput performance compared to conventional strategies at $T=50$; further, it works even though the channel simulation includes the computation error.
The proposed approach can help system operators design system parameters before deployment, for example, cell planning in 6G networks or access point deployment in public Wi-Fi.
\par
Note that the proposed framework utilizes the spatial correlation of throughput performances to search for optimal BS settings.
It depends on the spatial correlation of shadowing; thus, its performance improvement is limited under short correlation distance environments, including indoor scenarios with super-high-frequency (SHF) systems.
Furthermore, this study assumed isotropic antennas and static transmission positions for the BSs.
Introducing other adjustable parameters, such as beamformer and path planning in unmanned aerial vehicle (UAV)-based BSs, will allow further throughput; however, it increases the input dimensionality in the BO.
The BO design for more parameters remains a topic for future work.

\ifCLASSOPTIONcaptionsoff
  \newpage
\fi

\vspace{-1mm}
\bibliographystyle{IEEEbib}
\bibliography{refs}

\end{document}

%% file: introduction.tex
\section{Introduction}
Designing wireless systems before deployment is a typical, but essential task for efficient spectrum utilization. For example, in a sixth-generation (6G) network\cite{lee_ojvt2021}, operators must configure the parameters for dense base stations (BSs), such as BS placement and transmission power values, to provide stable wireless service in an area of interest under limited frequency resources.
Furthermore, public Wi-Fi\cite{yu-ieeeacm2017} and smart stadiums \cite{wu-ieeejsyst2022} are examples of such systems.
\par
If the wireless channel is influenced solely by path loss, then BS placement can be optimized using lightweight channel simulations based on empirical path loss models. However, in practice, shadowing effects often degrade the system performance. The shadowing effect, which is a stochastic process, can only be quantified after determining the positions of the transmitters through channel simulations or actual measurements, rendering white-box optimization frameworks impractical.
To proactively incorporate shadowing effects into the system design, it is necessary to iterate through parameter selection, channel simulation, and performance evaluation. However, high-precision channel simulations (e.g., ray tracing) are computationally intensive and often require several hours to days per evaluation, thereby limiting the feasibility of conducting numerous simulations.
This challenge herein motivates us to explore {\it how to find the optimal BS configurations with as few channel simulations as possible.}
\par
Recently, adaptive experimental design has gained traction across various fields, including robotics, neuroscience, and materials discovery\cite{greenhill_access2020}.
This data-driven framework designs experiments by iteratively sampling parameters, evaluating the performance, and predicting the performance for unknown inputs based on data-driven models.
By employing a Gaussian process (GP)-based black-box optimization method known as Bayesian optimization (BO), optimal parameters can be identified with a limited number of simulations, greatly reducing design costs; BO will enable high-performance BS configurations with reduced simulation costs.
\par
In this letter, we propose an adaptive experimental design framework for a simulation-based BS design that supports the joint optimization of placement and transmission power.
Some recent studies have shown the benefits of BO in wireless systems (e.g., task offloading in mobile edge computing\cite{jia_twc2024} and wireless resource allocation\cite{zhang_tccn2023}).
These studies suggest that BO-based optimizers can enhance system performance with fixed BS placements over time-varying wireless channels, even when performance metrics are treated as black boxes.
In contrast to these studies, our approach focuses on exploring BS placements and transmission power settings to achieve high throughput over spatially-varying channels \textit{before system deployment}.
The objective is to maximize the average throughput in an area of interest, modeled as a black-box function due to shadowing. We propose an optimizer based on a nested BO architecture, and numerical simulations demonstrate that our framework achieves superior average throughput per unit area with fewer channel simulations.

%% file: systemmodel.tex
\section{System Model}
\label{sect:systemmodel}
\begin{figure}[t]
   \centering
   \includegraphics[width=0.8\linewidth]{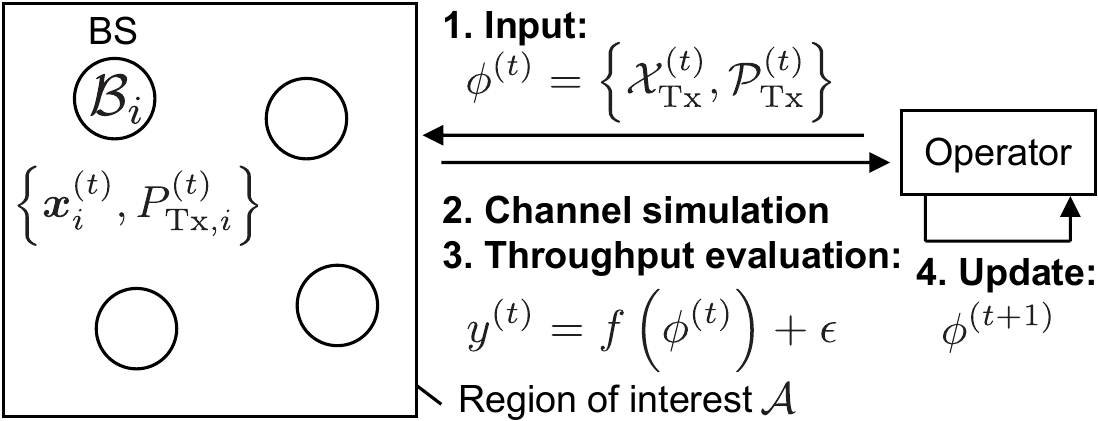}
   \caption{System model.}
\label{fig:systemmodel}
\end{figure}

We consider a scenario in which multiple BSs provide wireless communication services in a two-dimensional area.
Their operator tunes the BS placements and transmission power values before system deployment through several channel simulation trials to ensure that the average throughput within the area of interest is maximized, as shown in Fig.\,\ref{fig:systemmodel}.

\par
We define the set of BSs as $\mathcal{B}=\{\mathcal{B}_i \mid i=1, 2, \cdots, N_\mathrm{Tx}\}$.
where $\mathcal{B}_i$ denotes the $i$-th BS and $N_\mathrm{Tx}$ is the number of BSs. 
We also express the region of interest as $\mathcal{A}= \bigcup_{i=1}^{N_\mathrm{Tx}} \mathcal{A}_i$, where $\mathcal{A}_i (\in \mathbb{R}^2)$ is the possible coordinate of the $i$-th BS.
Assuming that the channel follows path loss and shadowing, the received signal power from the $i$-th coordinate ${\bm x}_{\mathrm{Tx},i} (\in \mathcal{A}_i)$ to the coordinate ${\bm x}(\in \mathcal{A})$ can be modeled as
\begin{equation}
    P_{\mathrm{Rx},i}({\bm x})=P_{\mathrm{Tx},i} d^{-\eta}_i w({\bm x}, {\bm x}_{\mathrm{Tx},i})\;\;\;\text{[mW]},
    \label{eq:channel-model}
 \end{equation}
 where $d_i=||{\bm x}_{\mathrm{Tx},i} - {\bm x}||$\,[m] is the communication distance ($||\cdot||$ is the Euclidean distance), $\eta$ is the path loss index, and $P_{\mathrm{Tx},i}$\,[mW] is the transmission power of the $i$-th BS.
 Furthermore, $w({\bm x}, {\bm x}_{\mathrm{Tx},i})$ represents the shadowing, indicating a spatial correlation between peer-to-peer wireless links\cite{wang-tvt2008}.
 \par
 When all BSs transmit signals simultaneously over a shared frequency band, the channel capacity of the $i$-th BS at ${\bm x}$ is given by
 \begin{equation}
    C_i({\bm x}) = B\log_2\left(1+\frac{P_{\mathrm{Rx},i}({\bm x})}{\sum_{\mathcal{B}\backslash \mathcal{B}_i}P_{\mathrm{Rx},j}({\bm x})+BN_0}\right)\;\;\text{[bps]}
    \label{eq:channel-capacity}
 \end{equation}
 where $B$\,[Hz] is the bandwidth and $N_0$\,[mW/Hz] is the additive white Gaussian noise (AWGN).
 Herein, we define a set of input parameters as $\phi=\{\mathcal{X}_\mathrm{Tx}, \mathcal{P}_\mathrm{Tx}\}$ where
 \begin{align}
    \mathcal{X}_\mathrm{Tx}&=\left\{{\bm x}_{\mathrm{Tx},i} \mid i=1, 2, \cdots N_\mathrm{Tx}\right\},\\
    \mathcal{P}_\mathrm{Tx}&=\left\{P_{\mathrm{Tx},i} \mid i=1, 2, \cdots N_\mathrm{Tx}\right\}.
\end{align}
\par
The objective is to maximize the average throughput per unit area $\left[\text{bps}/\text{m}^2\right]$ in the region of interest $\mathcal{A}$; i.e.,
\begin{maxi!}
    {\phi}
    {\left[ f(\phi) \coloneqq \frac{1}{|\mathcal{A}|}\sum_{i=1}^{N_\mathrm{Tx}}\int_\mathcal{A} C_i({\bm x})d{\bm x} \right]\label{eq:optimization-problem}}
    {\label{prb}} 
    {}
    \addConstraint{0\leq P_{\mathrm{Tx},i} \leq P_\mathrm{max},\;\forall i\label{subeq:constraint-power}}
    \addConstraint{{\bm x}_{\mathrm{Tx},i}\in \mathcal{A}_i, \forall i,\label{subeq:constraint-location}}
\end{maxi!}
 where $P_\mathrm{max}$\,[mW] is the maximum transmission power.
 \par
 Note that if the system can utilize multiple channels, the objective function can be extended to $f'(\phi) = \sum_{l=1}^{n_f} f_l(\phi)$ because of the orthogonality between channels, where $n_f$ denotes the number of channels and $f_l$ is the average throughput per unit area in the $l$-th channel. 
 Based on the additive decomposition of the objective function\cite{pmlr-v84-rolland18a}, the proposed method can support this by optimizing the placement and transmission powers for each $f_i$ by using the proposed framework.
 Since this extension is straightforward, this study focuses on improving the efficiency of a single channel.
 \par
 Furthermore, the proposed method can be extended to a 3D environment without modification by redefining the 2D vector ${\bm x}_{\mathrm{Tx},i}$ as a 3D vector. However, this extension increases the number of parameters that must be tuned, which may slightly degrade the convergence performance with respect to the number of simulations.

%% file: bo.tex
\section{Adaptive Experimental Design via BO}
\label{sect:background_bo}
Before introducing the proposed framework, this section summarizes the BO.
To present the BO separately from problem \eqref{prb}, we denote the set of input parameters to be optimized and the objective function as $\psi$ and $h(\psi)$, respectively. The objective in the BO can be given by 
\begin{equation}
    \psi_\mathrm{opt} = \underset{\psi \in \Psi_\mathrm{FR}} {\operatorname{argmax}}\;h(\psi),
\end{equation}
where $\Psi_\mathrm{FR}$ denotes the feasible region for the input parameters.
The method can be divided into three steps: (i) initialization, (ii) Gaussian process regression (GPR), and (iii) parameter update based on acquisition function.
We detail these steps as follows. This algorithm is summarized in Alg.\,\ref{alg:straightforward}.

\subsubsection{Initialization}
Initially, several simulation trials are performed to develop an initial dataset
\begin{equation}
  \mathcal{D}^{(0)} =\left\{\left[\psi^{(t)}, y^{(t)}\right]\relmiddle| t=1, 2, \cdots, N_\mathrm{init}\right\},
  \label{eq:initial-dataset}
\end{equation}
where $N_\mathrm{init}$ indicates the number of initial trials and $\psi^{(t)}$ indicates the $t$-th initial input randomly selected from the feasible region.
In addition, $y^{(t)} = h\left(\psi^{(t)}\right) + \epsilon$ is the observation value and $\epsilon$ is the observation noise.

\subsubsection{GPR}
This step estimates the distribution of the observation $p\left(y\relmiddle|\mathcal{D}^{(t)}\right)$ based on GPR, a kernel-based non-parametric regression method that assumes a GP for the target function.
To this end, we first tune a set of hyperparameters $\theta = \left\{{\bm \omega}, \sigma_\epsilon \right\}$, where ${\bm \omega}$ represents the hyperparameter vector for a kernel function $k$.
Note that $k$ measures the similarity between data points in a multi-dimensional space (e.g., the radial basis function (RBF))
The tuning of $\theta$ can be achieved by maximizing the following log-marginal likelihood:
\begin{align}
  \log p\left({\bm y}\relmiddle| {\bm X}, \theta\right) = &-\frac{1}{2}({\bm y}-{\bm m})^\mathrm{T}\left({\bm K} + \sigma^2_\epsilon {\bm I}\right)^{-1}({\bm y}-{\bm m}) \nonumber\\
 & - \frac{1}{2}\log\mathrm{det}\left({\bm K} + \sigma^2_\epsilon {\bm I}\right) -\frac{|\mathcal{D}|}{2}\log2\pi,
   \label{eq:log-marginal}
\end{align}
where ${\bm I}$ is the $|\mathcal{D}| \times |\mathcal{D}|$ identity matrix and ${\bm K}\in \mathbb{R}^{|\mathcal{D}| \times |\mathcal{D}|}$ is the kernel matrix, where its element is $K_{ij} = k(\psi^{(i)}, \psi^{(j)})$.
Further, ${\bm m}$ is a vector with $(t+N_\mathrm{init})$ elements, where its $i$-th element $m\left(\psi^{(i)}\right)$ represents the prior mean at $\psi^{(i)}$; it can be estimated as $m\left(\psi^{(i)}\right)= \frac{1}{|\mathcal{D}|}\sum_{t=1}^{|\mathcal{D}|}y^{(t)}, \forall i$.
\par
After the maximum likelihood estimation (MLE), the GPR predicts the distribution of the output for unknown inputs $\Psi_\ast = \left\{\psi_{\ast,i}\relmiddle| i=1, 2, \cdots,N_\mathrm{test}\right\}$,
 where $N_\mathrm{test}$ indicates the number of unknown inputs that must be interpolated.
For an unknown input $\psi_\ast$, the output mean and variance can be expressed by the following equations:

\begin{align}
  \mu (\psi_\ast) &=  m(\psi_\ast) + {\bm k}^{\mathrm{T}}(\Psi, \psi_\ast)\left({\bm K} + \sigma^2_\epsilon {\bm I}\right)^{-1}({\bm y}-{\bm m}) \label{eq:fullgpr-mean}\\
  \sigma^2 (\psi_\ast) &=  k(\psi_\ast, \psi_\ast) - {\bm k}^{\mathrm{T}}(\Psi, \psi_\ast)\left({\bm K} + \sigma^2_\epsilon {\bm I}\right)^{-1} {\bm k}(\Psi, \psi_\ast), \label{eq:fullgpr-var}
\end{align}
where $\Psi = \left\{\psi_i \relmiddle| i=1, 2, \cdots, t+N_\mathrm{init}\right\}$, and
\begin{equation}
  {\bm k}(\Psi, \psi_\ast) = \left[k(\psi_1, \psi_\ast), \cdots, k(t+N_\mathrm{init}, \psi_\ast)\right]^\mathrm{T}.
\end{equation}
Finally, we can predict the observation distribution at $\psi_\ast$ as the normal distribution with mean $\mu (\psi_\ast)$ and variance $\sigma^2 (\psi_\ast)$.

\subsubsection{Parameter Update}
Subsequently, this step estimates the acquisition function, which is a criterion for evaluating the goodness of the input parameters.
We employed expected improvement (EI), which quantifies the anticipated performance gain from sampling a specific point by considering both the potential for discovering a better solution and the uncertainty in the model's predictions; i.e.,
\begin{equation} 
  \alpha(\psi) = \mathbb{E}\left[\max\left(y-y^{+},0\right)\right],
  \label{eq:expected_improvement}
\end{equation}
where $y^{+}$ denotes the maximum value of the dataset.
Finally, the next input can be determined by
\begin{equation}
  \psi^{(t+1)} = \underset{\psi\in \Psi_\ast} {\operatorname{argmax}}\;\alpha(\psi).
  \label{eq:next_input_purebo}
\end{equation}
Steps 2 and 3 are repeated $T$ times; the optimized input $\psi_\mathrm{opt}$ is $\psi^{(t)} \in \mathcal{D}^{(T)}$ which maximizes $y^{(t)}$; that is,
\begin{align}
    \psi_\mathrm{opt} = \psi^{(t_\mathrm{best})},\,\,
    t_\mathrm{best} = \underset{t \in \{1, 2, \cdots, |\mathcal{D}^{(T)}|\}} {\operatorname{argmax}}\;y^{(t)}.
    \label{eq:optimized_input_purebo}
\end{align}

\begin{algorithm}[t]
    \caption{BO for adaptive experimental design}
    \label{alg:straightforward}
    \begin{algorithmic}[1]
        \REQUIRE $\mathcal{D}^{(0)} =\left\{\left[\psi^{(t)}, y^{(t)}\right]\relmiddle| t=1, 2, \cdots, N_\mathrm{init}\right\}$

        \FOR{$t=1, 2, \cdots, T$}
        \STATE Kernel tuning with MLE and $\mathcal{D}^{(t)}$
           \STATE Randomly sample a set $\Psi_\ast$
          \STATE Calculate $\mu$ and $\sigma^2$ for $\psi_{\ast,i} \in \Psi_\ast$ by Eqs.\,\eqref{eq:fullgpr-mean}\eqref{eq:fullgpr-var}
          \STATE Calculate $\alpha$ for $\psi_{\ast,i} \in \Psi_\ast$ by Eq.\,\eqref{eq:expected_improvement}
          \STATE Select $\psi^{(t+1)}$ by Eq.\,\eqref{eq:next_input_purebo}.
          \STATE Observe $y^{(t+1)} = h\left(\psi^{(t+1)}\right)+\epsilon$
          \STATE Update the dataset as $\mathcal{D}^{(t+1)} \!=\! \mathcal{D}^{(t)}\cup \left\{\left[\psi^{(t+1)}, y^{(t+1)}\right]\right\}$
        \ENDFOR
      \RETURN{$\psi_\mathrm{opt}$ calculated by Eq.\,\eqref{eq:optimized_input_purebo}}
        
    \end{algorithmic}
\end{algorithm}

%% file: bo_proposed.tex
\section{Applying BO to Wireless System Design}
\label{sect:proposedmethod}
BO models the set of observation values as a GP and interpolates the values at unobserved regions to identify better inputs.
In practice, $P_{\mathrm{Rx},i}$ follows the GP over the dBm domain spatially\cite{wang-tvt2008}, suggesting that $C_i$ can be approximated as GP at high signal-to-interference-plus-noise ratio (SINR).
Considering this background, we apply the BO algorithm to solve problem\,\eqref{prb}.
\par
\begin{figure}[t]
    \centering
    \subfigure[Alg.\,\ref{alg:straightforward}.]{\includegraphics[height=20mm]{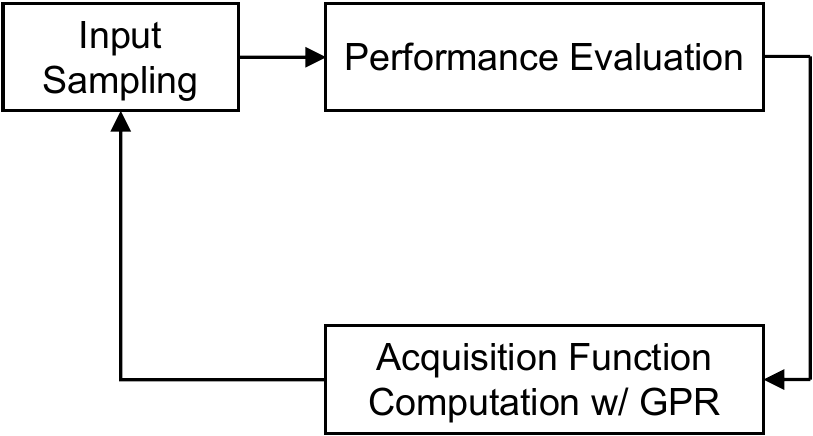}
    \label{subfig:bo_straightforward}}
    \subfigure[BO for problem \eqref{prb}.]{\includegraphics[height=20mm]{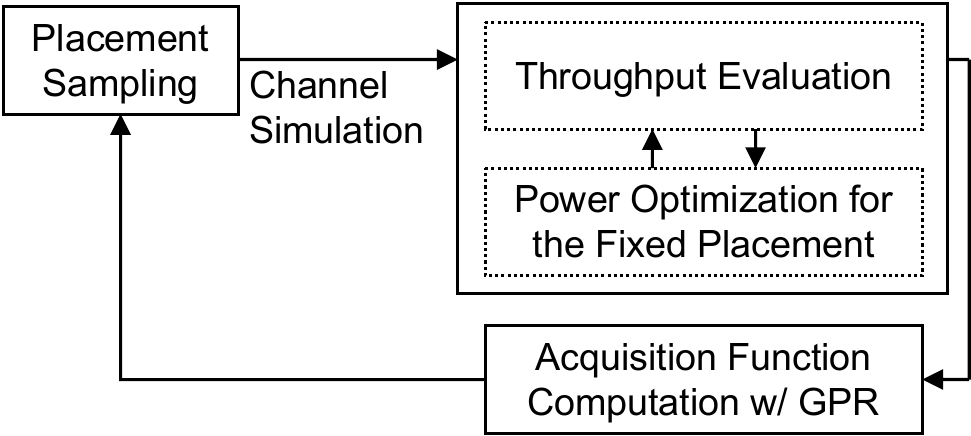}
    \label{subfig:bo_proposed}}
  \caption{Workflows in the pure BO and proposed framework.}\label{fig:bo_comparison}
\end{figure}

Although Alg.\,\ref{alg:straightforward} can simultaneously search for both $\mathcal{X}_\mathrm{Tx}$ and $\mathcal{P}_\mathrm{Tx}$, this straightforward implementation would suffer from the curse of dimensionality as $N_\mathrm{Tx}$ increases.
To avoid this problem, our framework introduces a nested BO architecture that separates the input parameters into two sets (Fig.\,\ref{fig:bo_comparison}).
This approach focuses on the fact that the transmission power values $\mathcal{P}_\mathrm{Tx}$ can be optimized using a one-channel simulation for a given $\mathcal{X}_\mathrm{Tx}$, which can improve the throughput performance in the BO without increasing the number of channel simulations.
\par
As with pure BO, our framework is divided into three steps, as outlined below.
This algorithm is summarized in Alg.\,\ref{alg:dimension-reduced}.
\subsubsection{Initialization}
At the beginning of the $t$-th step, this method randomly samples a set of transmitter placements $\mathcal{X}^{(t)}_\mathrm{Tx}$ and evaluates throughput performance.
It then optimizes the transmission power values such that
\begin{equation}
  \mathcal{P}_\mathrm{Tx, opt}^{(t)} = \underset{\mathcal{P}_\mathrm{Tx}} {\operatorname{argmax}}\;f\left(\mathcal{P}_\mathrm{Tx} \relmiddle| \mathcal{X}^{(t)}_\mathrm{Tx}\right).
  \label{eq:ptxopt_t_proposed}
\end{equation}
This result can be obtained by Alg.\,\ref{alg:straightforward} assuming that $\mathcal{X}^{(t)}_\mathrm{Tx}$ is fixed.
\par
In this framework, throughput performance $y^{(t)}$ can be obtained from a channel simulation.
Assuming that the channel simulation result for the $i$-th transmitter can express the average channel gain at a pair of transmitter and receiver coordinates, its logarithmic form can be expressed as a function

\begin{equation}
  g^{(t)}_{i}\left({\bm x}\right) = -10\eta\mathrm{log}_\mathrm{10}\left|\left|{\bm x}^{(t)}_\mathrm{Tx, i} - {\bm x}\right|\right| + 10\mathrm{log}_\mathrm{10} w\left({\bm x}, {\bm x}^{(t)}_\mathrm{Tx, i}\right) + \epsilon_\mathrm{M}
  \label{eq:ckm-model}
\end{equation}
where $\epsilon_\mathrm{M}$ is the simulation error and ${\bm x}\in \mathcal{A}$; after the simulation is performed for all possible transmitters. Accordingly, a set of channel simulation results can be constructed as
\begin{equation}
    \mathcal{G}^{(t)}=\left\{g^{(t)}_{i}({\bm x})\relmiddle| i=1, 2, \cdots, N_\mathrm{Tx}\right\}.
\end{equation}
Then, we obtain throughput $y^{(t)}$ using $\mathcal{P}^{(t)}_\mathrm{Tx, opt}$ and $\mathcal{G}^{(t)}$, and Eq.\,\eqref{eq:optimization-problem}.
Once $y^{(t)}$ is observed, the dataset is updated as

\begin{equation}
  \mathcal{D}^{(t)} = \mathcal{D}^{(t-1)}\bigcup \left\{\phi^{(t)}, y^{(t)}\right\},
  \label{eq:updated_dataset1_proposed}
\end{equation}
where $\phi^{(t)} = \left\{\mathcal{X}_\mathrm{Tx}^{(t)}, \mathcal{P}_\mathrm{Tx, opt}^{(t)}\right\}$.
These steps are repeated $N_\mathrm{init}$ times for the initial dataset.
\subsubsection{GPR}
To tune the set of placements, this step performs (a)\,MLE with $\mathcal{D}^{(t)}$ as with simultaneous BO, and (b)\,GPR over the set of placements $\mathcal{X}_\mathrm{Tx}$.
The GPR is performed for a set of unknown inputs $\mathcal{X}_{\ast} = \left\{\mathcal{X}_{\ast,i}\relmiddle| i=1, 2, \cdots, N_\mathrm{test}\right\}$, and $p\left(y\relmiddle| \mathcal{D}^{(t)}\right)$ is modeled for $\mathcal{X}_\mathrm{Tx} \in \mathcal{X}_{\ast}$.
As with Eqs.\,\eqref{eq:fullgpr-mean}\eqref{eq:fullgpr-var}, the mean and variance of the output at $\mathcal{X}_{\ast,i}$ can be given by
\begin{align}
  \mu (\mathcal{X}_{\ast,i}) &=  m(\mathcal{X}_{\ast,i}) + {\bm k}^{\mathrm{T}}_{\ast,i}\left({\bm K} + \sigma^2_\epsilon {\bm I}\right)^{-1}({\bm y}-{\bm m})\label{eq:mean_gpr_proposed}\\
  \sigma^2 (\mathcal{X}_{\ast,i}) &= k(\mathcal{X}_{\ast,i}, \mathcal{X}_{\ast,i}) - {\bm k}^{\mathrm{T}}_{\ast,i}\left({\bm K} + \sigma^2_\epsilon {\bm I}\right)^{-1} {\bm k}_{\ast,i},\label{eq:var_gpr_proposed}
\end{align}
where ${\bm k}_{\ast,i} = k(\xi, \mathcal{X}_{\ast,i})$, and
\begin{align}
  \xi = \left\{\mathcal{X}^{(i)}_{\mathrm{Tx}} \relmiddle| i=1, 2, \cdots, t+N_\mathrm{init}\right\}.
\end{align}

\subsubsection{Placement Update and Power Optimization}
This step calculates the acquisition function $\alpha(\mathcal{X}_\mathrm{Tx})$ for $\mathcal{X}_\mathrm{Tx} \in \mathcal{X}_{\ast}$ based on EI.
The next set of placements can be then selected by
\begin{equation}
  \mathcal{X}_\mathrm{Tx}^{(t+1)} = \underset{\mathcal{X}_\mathrm{Tx} \in \mathcal{X}_{\ast}} {\operatorname{argmax}}\;\alpha(\mathcal{X}_\mathrm{Tx}).
  \label{eq:next_placement_proposed}
\end{equation}
After the channel information $\mathcal{G}^{(t+1)}$ is simulated, based on Alg.\,\ref{alg:straightforward}, we optimize the transmission power values such that 
\begin{equation}
  \mathcal{P}_\mathrm{Tx, opt}^{(t+1)} = \underset{\mathcal{P}_\mathrm{Tx}} {\operatorname{argmax}}\;f\left(\mathcal{P}_\mathrm{Tx} \relmiddle| \mathcal{X}^{(t+1)}_\mathrm{Tx}, \mathcal{G}^{(t+1)}\right).
  \label{eq:ptxopt_tplus1_proposed}
\end{equation}
Then, the dataset is updated to
\begin{equation}
  \mathcal{D}^{(t+1)} = \mathcal{D}^{(t)}\bigcup \left\{\phi^{(t+1)}, y^{(t+1)}\right\}.
  \label{eq:updated_dataset2_proposed}
\end{equation}
Steps 2 and 3 are repeated $T$ times.
Finally, the optimized input $\phi_\mathrm{opt}$ is $\phi^{(t)} \in \mathcal{D}^{(T)}$ that maximizes $y^{(t)}$: this operation can be given by,
\begin{equation}
    \phi_\mathrm{opt} = \phi^{(t_\mathrm{best})},\,\,
    t_\mathrm{best} = \underset{t \in \{1, 2, \cdots, |\mathcal{D}^{(T)}|\}} {\operatorname{argmax}}\;y^{(t)}.
    \label{eq:optimized_input_proposed}
\end{equation}

\begin{algorithm}[t]
    \caption{BO for channel simulation-based BS design}
    \label{alg:dimension-reduced}
    \begin{algorithmic}[1]
    \REQUIRE $\mathcal{D}^{(0)} = \emptyset$

    \FOR{$t=1, 2, \cdots, N_\mathrm{init}$}
      \STATE Randomly select the set of placements $\mathcal{X}_\mathrm{Tx}^{(t)}$
      \STATE Construct $\mathcal{G}^{(t)}$ by channel simulation.
      \STATE Find $\mathcal{P}_\mathrm{Tx, opt}^{(t)}$ by Eq.\,\eqref{eq:ptxopt_t_proposed} and Alg.\,\ref{alg:straightforward}
      \STATE Set $\phi^{(t)} = \left\{\mathcal{X}_\mathrm{Tx}^{(t)}, \mathcal{P}_\mathrm{Tx, opt}^{(t)}\right\}$
        \STATE Observe throughput $y^{(t)}$ by $\mathcal{P}_\mathrm{Tx, opt}^{(t)}$ and $\mathcal{G}^{(t)}$.
      \STATE Update $\mathcal{D}^{(t-1)}$ to $\mathcal{D}^{(t)}$ by Eq.\,\eqref{eq:updated_dataset1_proposed}
    \ENDFOR
    \FOR{$t=1, 2, \cdots, T$}
      \STATE Kernel tuning with MLE and $\mathcal{D}^{(t)}$
      \STATE Sample $\mathcal{X}_{\ast} = \left\{\mathcal{X}_{\ast,i}\relmiddle| i=1, 2, \cdots, N_\mathrm{test}\right\}$
      \STATE Calculate $\mu$ and $\sigma^2$ for $\mathcal{X}_{\ast,i} \in \mathcal{X}_{\ast}$ by Eqs.\,\eqref{eq:mean_gpr_proposed}\eqref{eq:var_gpr_proposed}
      \STATE Calculate $\alpha$ for $\mathcal{X}_{\ast,i} \in \mathcal{X}_{\ast}$ by Eq.\,\eqref{eq:expected_improvement}
      \STATE Select $\mathcal{X}_\mathrm{Tx}^{(t+1)}$ by Eq.\,\eqref{eq:next_placement_proposed}
      \STATE Simulate $\mathcal{G}^{(t+1)}$ for $\mathcal{X}_\mathrm{Tx}^{(t+1)}$
      \STATE Find $\mathcal{P}_\mathrm{Tx, opt}^{(t+1)}$ by Eq.\,\eqref{eq:ptxopt_tplus1_proposed} and Alg.\,\ref{alg:straightforward}
      \STATE Observe throughput $y^{(t+1)}$ by $\mathcal{P}_\mathrm{Tx, opt}^{(t+1)}$ and $\mathcal{G}^{(t+1)}$.
      \STATE Update $\mathcal{D}^{(t)}$ to $\mathcal{D}^{(t+1)}$ by Eq.\,\eqref{eq:updated_dataset2_proposed}
    \ENDFOR
  
  \RETURN{$\phi_\mathrm{opt}$ calculated by Eq.\,\eqref{eq:optimized_input_proposed}}
    
    \end{algorithmic}
\end{algorithm}

%% file: performance.tex
\vspace{-1mm}
\section{Performance Evaluation}
\label{sect:performance}
The proposed framework was implemented using Optuna 3.2.0\cite{optuna}, BoTorch 0.9.2\cite{botorch}, and Python 3.10.6.
Furthermore, the performance of the proposed framework (Alg.\,\ref{alg:dimension-reduced}) was compared with that of the following three baseline methods.
\begin{itemize}
  \item \textbf{Placement Optimization}: Assuming the maximum transmission power, this method optimizes $\mathcal{X}_\mathrm{Tx}$ via Alg.\,\ref{alg:straightforward}.
  \item \textbf{Regular Placement with Power Optimization}: This method determines $\mathcal{X}_\mathrm{Tx}$ as a regular hexagonal grid. It then optimizes $\mathcal{P}_\mathrm{Tx}$ based on Alg.\,\ref{alg:straightforward}.
  \item \textbf{Pure BO}: This method simultaneously searches for $\mathcal{X}_\mathrm{Tx}$ and $\mathcal{P}_\mathrm{Tx}$ based on Alg.\,\ref{alg:straightforward} only.
  \item \textbf{Random Search}: This method samples the parameters in $\phi$ based on the uniform distribution.
\end{itemize}

The simulation parameters are listed in Table\,\ref{table:simulation-parameters}.
We simulated wireless channels based on Eq.\,\eqref{eq:channel-model}.
Shadowing exhibits spatial correlations that depend on the extent of change in the coordinates of the transmitter and receiver\cite{wang-tvt2008}.
To consider this effect, we assume that shadowing has a spatial correlation with the standard deviation $\sigma_\mathrm{S}$\,[dB].
Note that although it can be simulated by a multivariate log-normal distribution, generating it using a pure multivariate log-normal distribution in our simulation environment is computationally expensive.
Thus, we generated the shadowing based on the following:\,(i) allocating $N_\mathrm{s}$ points on $\mathcal{A}$ based on the Poisson point process (PPP), (ii)\ assigning i.i.d. log-normal random variables to the allocated points, and (iii) interpolating shadowing values on the missing coordinates based on the nearest neighborhood search.
The point density in PPP is designed such that the nearest neighbor distance $r$ satisfies $\mathbb{E}[r] = \mathbb{E}[d_\mathrm{cor}]$.
Note that we define the distance between a pair of wireless links as
$d_{ij} \triangleq \sqrt{||{\bm x}_{\mathrm{Tx},i} - {\bm x}_{\mathrm{Tx},j}||^2 + ||{\bm x}_{\mathrm{Rx},i} - {\bm x}_{\mathrm{Rx},j}||^2}$.

\begin{table}[t]
  \caption{Simulation parameters.}
  \vspace{-2mm}
  \label{table:simulation-parameters}
  \centering
   \begin{tabular}{p{2.2cm}|p{5.5cm}}
    \hline
     Parameter & Detail \\\hline
     Kernel function $k$ & RBF kernel without scaling factor\\
     Optimizer for MLE & L-BFGS-B\cite{L-BFGS-B} \\
     $N_\mathrm{init}$ and $N_\mathrm{test}$ & 8 and 256\\
     Size of $\mathcal{A}$ & $1000\,\mathrm{[m]} \times 1000\,\mathrm{[m]}$\\
     Range of $\mathcal{A}_i$ & $[0, 1000]$ ($x$ axis), $\left[\frac{1000}{N_\mathrm{Tx}}i, \frac{1000}{N_\mathrm{Tx}}(i+1)\right]$ ($y$ axis)\\
     Grid size & $20\,\mathrm{[m]} \times 20\,\mathrm{[m]}$\\
     $\eta$ & 4.0\\
     $d_\mathrm{cor}$ & 200\,[m]\\
     $\sigma_\mathrm{S}$ & 6.0\,[dB]\\
     $P_\mathrm{max}$ & 10\,[mW]\\
     $N_0$ & -174\,[dBm/Hz]\\
     $B$ & 20\,[MHz]\\
    \hline
   \end{tabular}
 \end{table}
\begin{figure}[t]
    \centering
    \subfigure[$N_\mathrm{Tx}=7$.]{\includegraphics[width=0.48\linewidth]{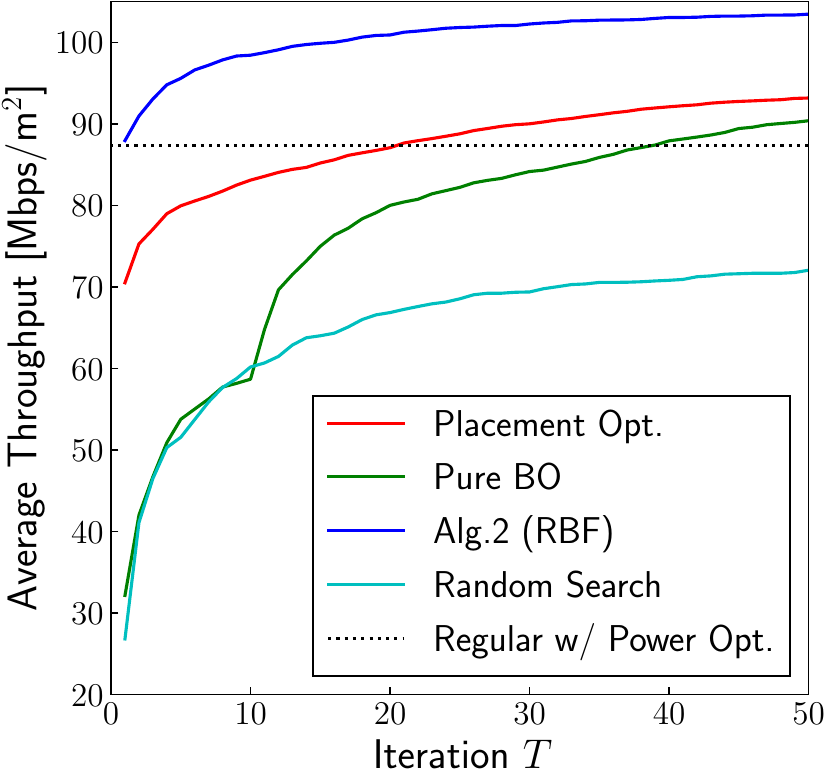}
    \label{subfig:effects_iteration_7tx}}
    \subfigure[$N_\mathrm{Tx}=19$.]{\includegraphics[width=0.48\linewidth]{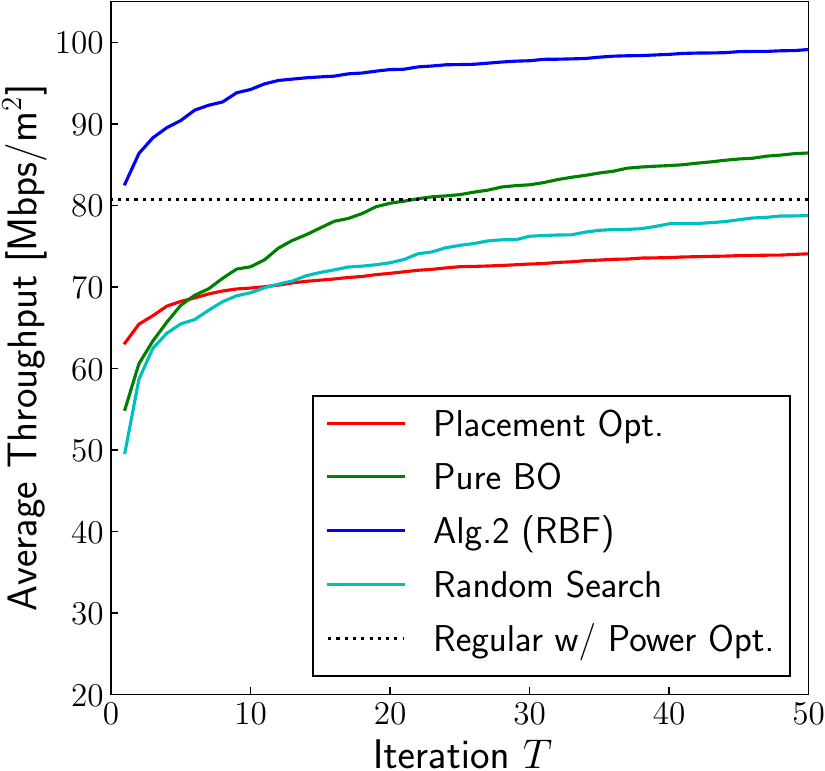}
    \label{subfig:effects_iteration_19tx}}
  \caption{Effects of the number of iterations.}\label{fig:effects_iteration}
\end{figure}

Fig.\,\ref{fig:effects_iteration} shows the effects of $T$ on the average throughput performance.
Assuming the errorless channel simulation, we iterated the independent simulations 100 times and calculated the average throughput for each condition.
The number of transmitters was set to $N_\mathrm{Tx}=7$\,(Fig.\,\ref{subfig:effects_iteration_7tx}) and $N_\mathrm{Tx}=19$\,(Fig.\,\ref{subfig:effects_iteration_19tx}).
For $N_\mathrm{Tx}=7$, the average throughput in the regular placement was 87.4 $\text{Mbps/m}^2$.
The other methods improved the throughput values as the number of iterations increased. At $T=50$, each method achieves the following throughput values: 93.2 $\text{Mbps/m}^2$ (placement optimization), 90.4 $\text{Mbps/m}^2$ (pure BO), 103.5 $\text{Mbps/m}^2$ (proposed method), and 72.1 $\text{Mbps/m}^2$ (random search).
Furthermore, placement optimization and pure BO outperformed the regular placement method at $T\geq 21$ and $T \geq 39$, respectively. The proposed framework resulted in a higher throughput than these methods. It outperformed regular placement, even with fewer iterations, and an 18.4\% improvement was confirmed at $T=50$. Moreover, the random search method is inferior to the other methods.
\par
Similar to the case of $N_\mathrm{Tx}=7$, an improvement in throughput with an increasing number of iterations was also observed for $N_\mathrm{Tx}=19$.
In contrast to the case of $N_\mathrm{Tx}=7$, the performance of the placement optimization was inferior to that of the regular placement method and random search for $N_\mathrm{Tx}=19$. For $N_\mathrm{Tx}=7$, the number of transmitters was small relative to the area size; a higher transmission power tends to improve the throughput performance in this case. However, as $N_\mathrm{Tx}$ increases, the interference between the transmitters becomes more significant, and the throughput deteriorates.
Pure BO and the proposed framework outperformed the other methods. Specifically, our framework improved throughput by 22.8\% compared to the regular placement method at $T=50$.
\par
Overall, for both $N_\mathrm{Tx}=7$ and $19$, the proposed methods achieved a higher throughput performance with a limited number of simulations compared with the baseline methods.

\begin{figure}[t]
    \centering
    \subfigure[$N_\mathrm{Tx}=7$.]{\includegraphics[width=0.48\linewidth]{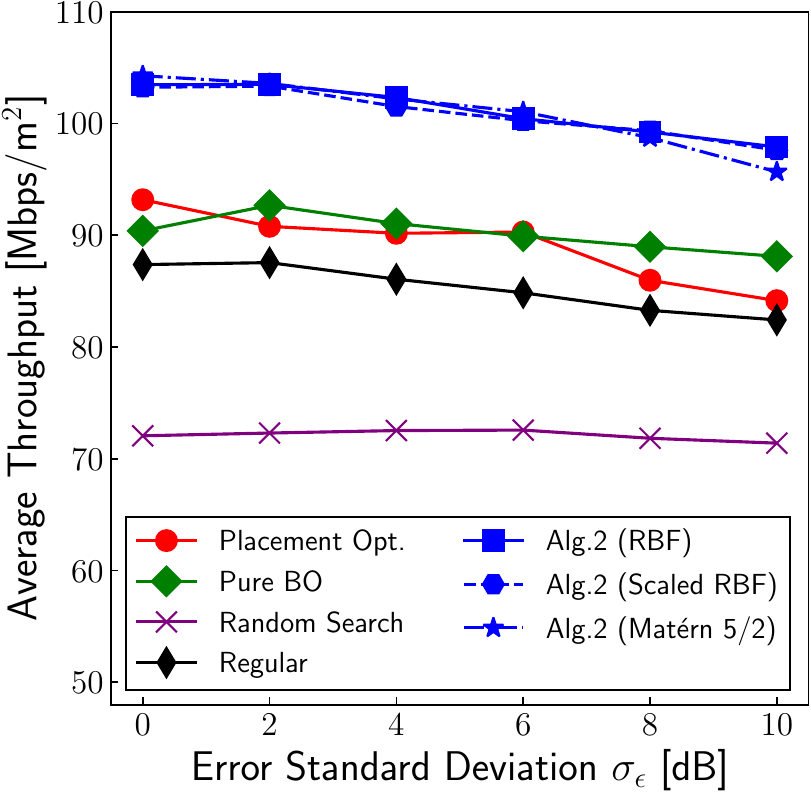}
    \label{subfig:effects_noise_7tx}}
    \subfigure[$N_\mathrm{Tx}=19$.]{\includegraphics[width=0.48\linewidth]{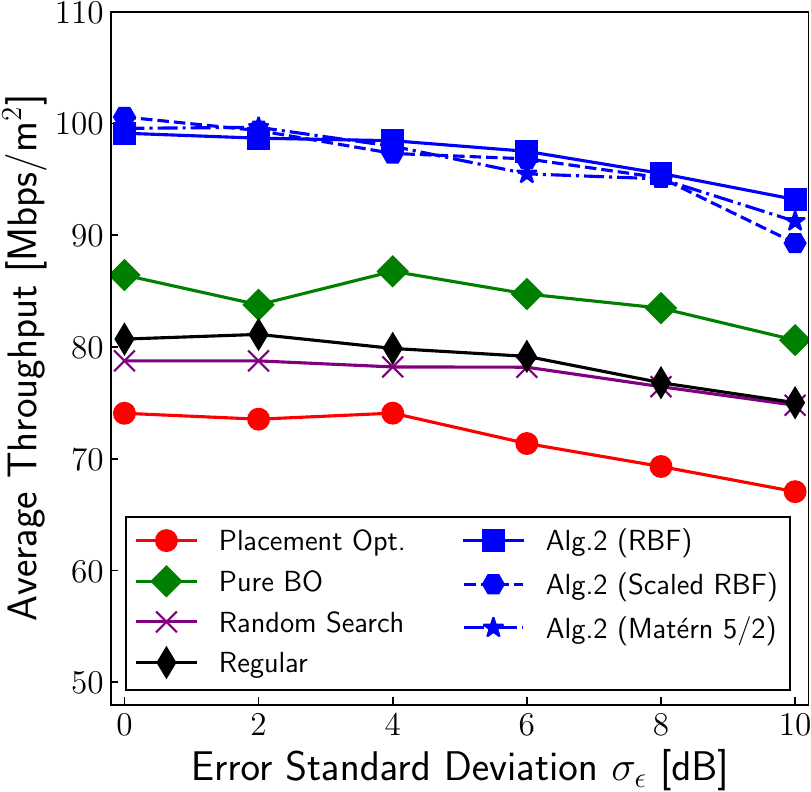}
    \label{subfig:effects_noise_19tx}}
  \caption{Effects of channel simulation error at $T=50$.}\label{fig:effects_noise}
  \vspace{-2mm}
\end{figure}

We then evaluated the effects of the observation error on the average throughput performance to discuss more practical performance\footnote{For example, based on \cite{mbugua-ieeeojap2021}, the RMSE of the ray-tracing simulation in sub-6\,GHz outdoor scenarios was approximately 4-13\,dB. Furthermore, it is necessary to average the samples that are within a half-wavelength radius around a certain reception point to mitigate the effects of multipath fading; however, this can cause several errors in shadowing estimation.}.
We model the error factor $\epsilon_\mathrm{M}$ in Eq.\,\eqref{eq:ckm-model} as a random variable following the normal distribution $\mathcal{N}(0, \sigma^2_\epsilon)$, where $\sigma_\epsilon$\,[dB] is the standard deviation.
Fig.\,\ref{fig:effects_noise} shows the effects of the standard deviation $\sigma_\epsilon$ on the average throughput performance where $N_\mathrm{Tx}=7$\,(Fig.\,\ref{subfig:effects_noise_7tx}) and $19$\,(Fig.\,\ref{subfig:effects_noise_19tx}).
We changed $\sigma_\epsilon$ from zero (noiseless) to 10 and evaluated the average throughput at $T=50$.
Furthermore, to confirm effects of kernel function, we also evaluated the proposed method with Matérn 5/2 kernel and scaled RBF kernel.

The throughput of all methods decreases as the noise standard deviation increases. For the proposed method (with the RBF kernel), the throughput decreased by 5.4\% at $N_\mathrm{Tx} = 7$ and by 6.0\% at $N_\mathrm{Tx} = 19$ compared to the noiseless case when $\sigma_\epsilon = 10$.
However, the proposed framework achieves better throughput than the baseline methods, as in the noiseless case.
For example, for $N_\mathrm{Tx}=7$, the throughput was approximately 18.8 \% higher than that of the regular placement at $\sigma_\epsilon=10$.
By modeling the observation with noise (i.e., $y^{(t)}=f\left(\phi^{(t)}\right) + \epsilon$) and performing black-box optimization while considering the uncertainties, BO facilitates error-robust transmission power and placement designs.
\par
Focusing on the impact of the kernel in the proposed method, similar trends were observed for all types of kernels.
In particular, at $\sigma_\epsilon = 10$, the RBF without scaling exhibited the best performance, achieving 103.5 $\text{Mbps/m}^2$ at $N_\mathrm{Tx} = 7$ and 99.1 $\text{Mbps/m}^2$ at $N_\mathrm{Tx} = 19$.
This kernel has fewer trainable parameters compared to other kernel functions.
When the relationship between samples can be sufficiently modeled with this kernel alone, it tends to exhibit better convergence properties in comparison to more complex kernels.
Therefore, the RBF without scaling is a reasonable choice for this problem.